\documentclass[letterpaper]{article} 
\usepackage[preprint]{aaai2027}  
\usepackage[hyphens]{url}  
\usepackage{graphicx} 
\usepackage{natbib}  
\usepackage{caption} 
\usepackage{algorithm}
\usepackage{algorithmic}

\usepackage{newfloat}
\usepackage{listings}
\DeclareCaptionStyle{ruled}{labelfont=normalfont,labelsep=colon,strut=off} 
\floatstyle{ruled}
\newfloat{listing}{tb}{lst}{}
\floatname{listing}{Listing}

\usepackage{booktabs}

\usepackage[utf8]{inputenc} 
\usepackage[T1]{fontenc}    
\usepackage{url}            
\usepackage{booktabs}       
\usepackage{amsfonts}       
\usepackage{nicefrac}       
\usepackage{microtype}      
\usepackage{xcolor}         

\usepackage{amsmath}
\usepackage{amssymb}
\usepackage{mathtools}
\usepackage{amsthm}
\usepackage{multirow}
\usepackage{pdfpages}
\usepackage{graphicx}
\usepackage[capitalize,noabbrev]{cleveref}
\theoremstyle{plain}
\newtheorem{theorem}{Theorem}[section]
\newtheorem{proposition}[theorem]{Proposition}

\theoremstyle{definition}

\theoremstyle{remark}

\usepackage{enumitem}
\usepackage{booktabs} 

\title{Semantic Tokens Beat IDs in Large Ranking Models: A Scaling Law Perspective}
\author{
    Zhen Zhao,
    Qile Zhang,
    Tong Zhang,
    Leyuan Yang, 
    Jie Xu, Qingliang Cai, \\
    Daorui Xiao \corresponding,
    Xiaojia Chang
}
\affiliations{
    \textsuperscript{\rm } ByteDance\\

    {zhaozhen.0718, zhangqile.111, zhangtong.z, xiaodaorui}@bytedance.com
}

\begin{document}

\maketitle

\begin{abstract}
Recent studies on scaling up ranking models have achieved substantial improvement for recommendation systems and search engines. However, most large-scale ranking systems rely on item and query IDs, where each item or query is treated as an independent categorical symbol and mapped to a learned embedding. As items and queries rapidly appear and disappear, these embeddings become difficult to train and maintain. This instability impedes effective learning of neural network parameters and limits the scalability of ranking models. In this paper, we show that semantic tokens possess greater scaling potential compared to item and query IDs. Our proposed framework TRM improves the token generation and application pipeline, leading to a 72\% reduction in sparse storage while achieving a 0.85\% increase in AUC. Extensive experiments further show that TRM consistently outperforms state-of-the-art models as model capacity scales.
Finally, TRM has been successfully deployed on large-scale personalized search engines, yielding a 0.26\% improvement in user active days and a 0.75\% reduction in change query ratio through an A/B test. We present the first complete replacement of both item and query IDs with semantic tokens in large-scale ranking scenarios, demonstrating superior scalability.
\end{abstract}


\section{Introduction}

Large Ranking Models (LRMs) serve as the backbone of modern recommendation systems \cite{zhang2024wukong,rankmixer,dhen} and search engines\cite{chen2025onesearch,search2}, playing a pivotal role in efficient information distributing.

Recently, scaling up model parameters in LRMs has become a promising approach for developing ranking models\cite{rankmixer,dhen,scaling1,scaling2}. Inspired by the success of scaling law in LLM\cite{kaplan2020scaling}, recent efforts focus on creating the most comprehensive model architecture for dense parameter scaling\cite{dhen,rankmixer}. However, the dynamic and unstable nature of categorical features of item IDs prevents efficient knowledge sharing: newly introduced IDs often suffer from cold-start problem, while old retired IDs discard all previously learned knowledge \cite{zhu2021mwuf}. Therefore, the rapid and dramatic shifting of the ID-based feature distribution hinders the learning of dense parameters, particularly in large-scale ranking systems \cite{zheng2025semantic}. As a result, the effectiveness and capability of scaling up dense parameters is compromised.

\begin{figure}
\centering
    \includegraphics[width=0.95\linewidth]{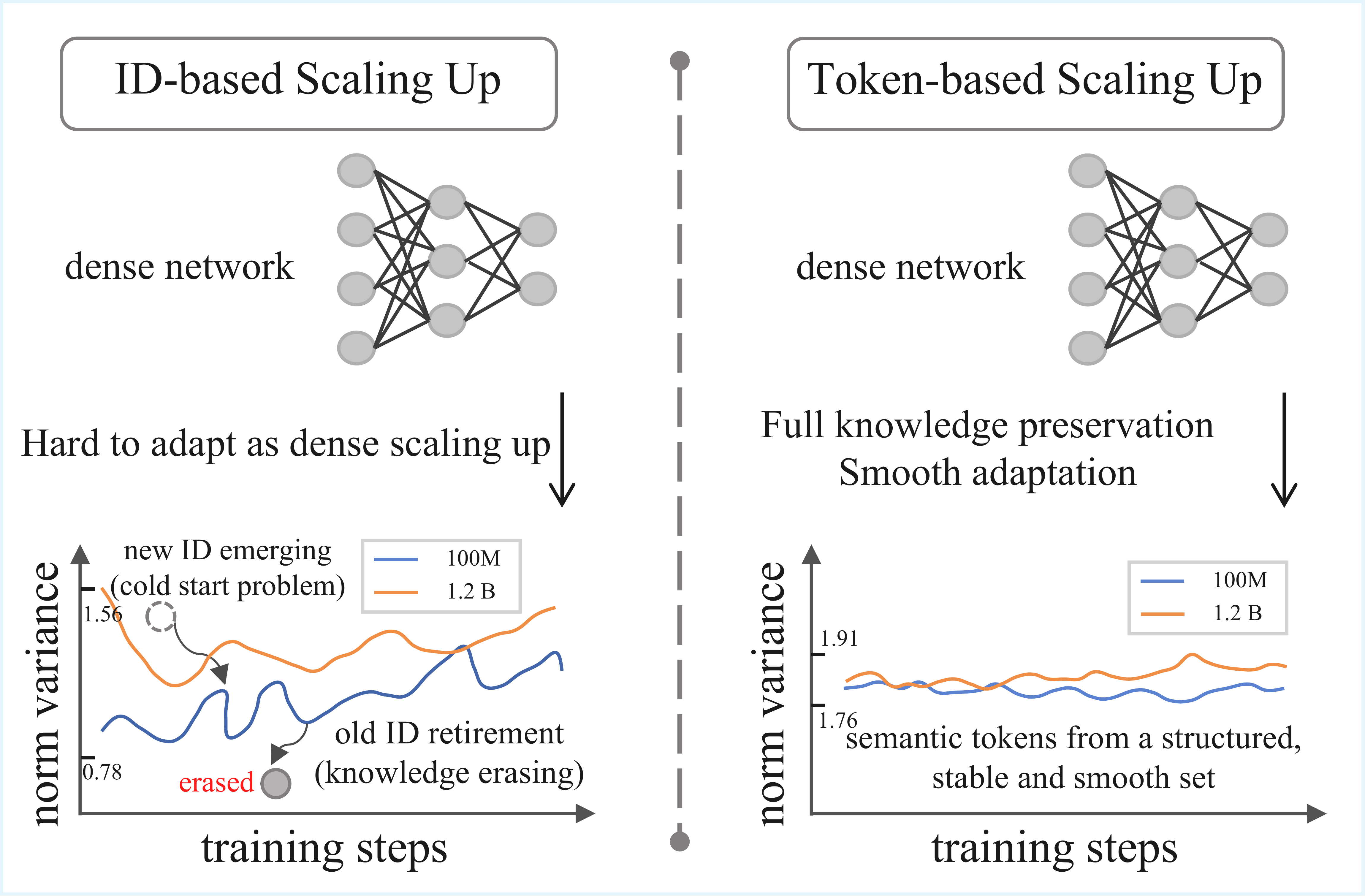}
    \caption{Illustration of different impacts of item IDs and semantic tokens to the scaling up of ranking models. Semantic tokens demonstrate more stable distribution as the number of parameter scales.}
    \label{fig:intro}
\end{figure}

In contrary, we propose to achieve scaling up dense parameters by replacing item IDs with semantic tokens\cite{rajput2023generative,zheng2025semid}. One of our key observations is that, semantic tokens construct a structural closed set, which is relatively more stable and smooth during training. As shown in Figure \ref{fig:intro}, we use norm variance \cite{norm-var} to measure the distribution change of ID embedding and token embedding. It is clear that semantic tokens have a more stable distribution during training as the number of parameter scales. 

Contradictory to our initial expectations, the experiments demonstrate that naively replacing item IDs with current semantic tokens in existing ranking models leads to immediate drop in performance, although this degradation diminishes as dense parameters increase. We provide three fundamental insights followed for this phenomenon. First, current semantic tokens typically incorporate their multi-modal information (image and text about the item, {\it etc}) for residual clustering, while ignoring the user-action domain that contains fundamentally different structural information from the vision/language/audio domain. Second, our experiments show that existing semantic tokens trade the memorization capability for better generalization performance due to coarse-grained clustering. Therefore, the ranking model fails to capture fine-grained combinative knowledge, leading to degraded performance for frequently occurring old items. Third, current approach directly combines the semantic tokens for input features, neglecting the structural information inside the token sequence.

To solve the above problems, we propose a novel token-based scaling up framework for ranking models, named TRM (Token-based Recommendation Model). Specifically, we develop a collaborative-filtering method to integrate user-action information into the original vision-language embedding model, enabling the semantic tokens clustered in both multi-modal and personalization domain. To address memorization issue, we propose independently learning each item's combinatorial knowledge, which can better balance between generalization and memorization of semantic tokens. Lastly, we developed a novel training framework that jointly optimizes the discriminative and the generative objectives. This framework not only leverages the accuracy of discriminative prediction but also exploits the structural information inside token sequences.

Experimentally, we evaluate TRM within a personalized search engine that relies on large-scale ranking models for online service. By applying TRM to state-of-the-art ranking models such as RankMixer and replacing both item and query IDs with semantic tokens, our approach achieves significant performance improvements and resource savings. Offline experiments on real-world user logs demonstrate a 0.65\% increase in AUC and a 72\% reduction in sparse storage. Furthermore, as the dense ranking model scales, TRM consistently outperforms both ID-based and other token-based models, with its relative QAUC gain increasing from 0.54\% to 0.85\%.

The main contribution of our work can be summarized as follows:
\begin{itemize}

\item We perform both theoretical and experimental analyses to prove the better scaling up performance of token-based models against ID-based models.

\item We developed the token-based scaling up framework TRM, which effectively solves the problem of user-action mis-alignment, memorization sacrifice and structural information missing of traditional semantic tokens, leading to a better scaling up performance.

\item  TRM has been successfully deployed on large-scale search engines, yielding 72\% sparse storage decrease, 0.26\% increase in user active days and 0.75\% decrease in change query ratio.
\end{itemize}

\section{Related Work}
\paragraph{ID-based large ranking models and the embedding bottleneck.}
Large-scale ranking models are traditionally dominated by sparse categorical features (e.g., item IDs) where each item is mapped to a learnable embedding in a large table, as exemplified by DLRM-style architectures \citep{naumov2019dlrm,rec-1,rec-2,rec-3,rec-4}. Under this paradigm new IDs continuously appear while old IDs retire. This feature drift problem complicates long-horizon knowledge accumulation and makes naive scaling by enlarging embedding tables inefficient \citep{coleman2023unified}.

\paragraph{Semantic IDs and discrete item tokenization.}
To mitigate the instability and poor generalization of raw IDs or hashed IDs, a growing line of work explores \emph{Semantic IDs}: compact discrete codes derived from content embeddings via vector quantization. \cite{singh2023semanticids} shows that semantic IDs can improve generalization for new and long-tail items in an industry-scale ranking model, and proposes sub-piece tokenization (including SentencePiece-style segmentation) to aid adaptation \citep{kudo2018sentencepiece}. In ads ranking, \cite{zhao2023embedding_survey} further analyze ID drifting and embedding instability, and \cite{zheng2025semid} propose prefix-based parameterizations to create semantically meaningful collisions. A recent comprehensive survey \citep{liu2024vq4rec} categorizes VQ methods for recommender systems into efficiency-oriented and quality-oriented approaches, highlighting their role in bridging content and collaborative signals. 

\begin{figure*}
    \centering
    \includegraphics[width=1\linewidth]{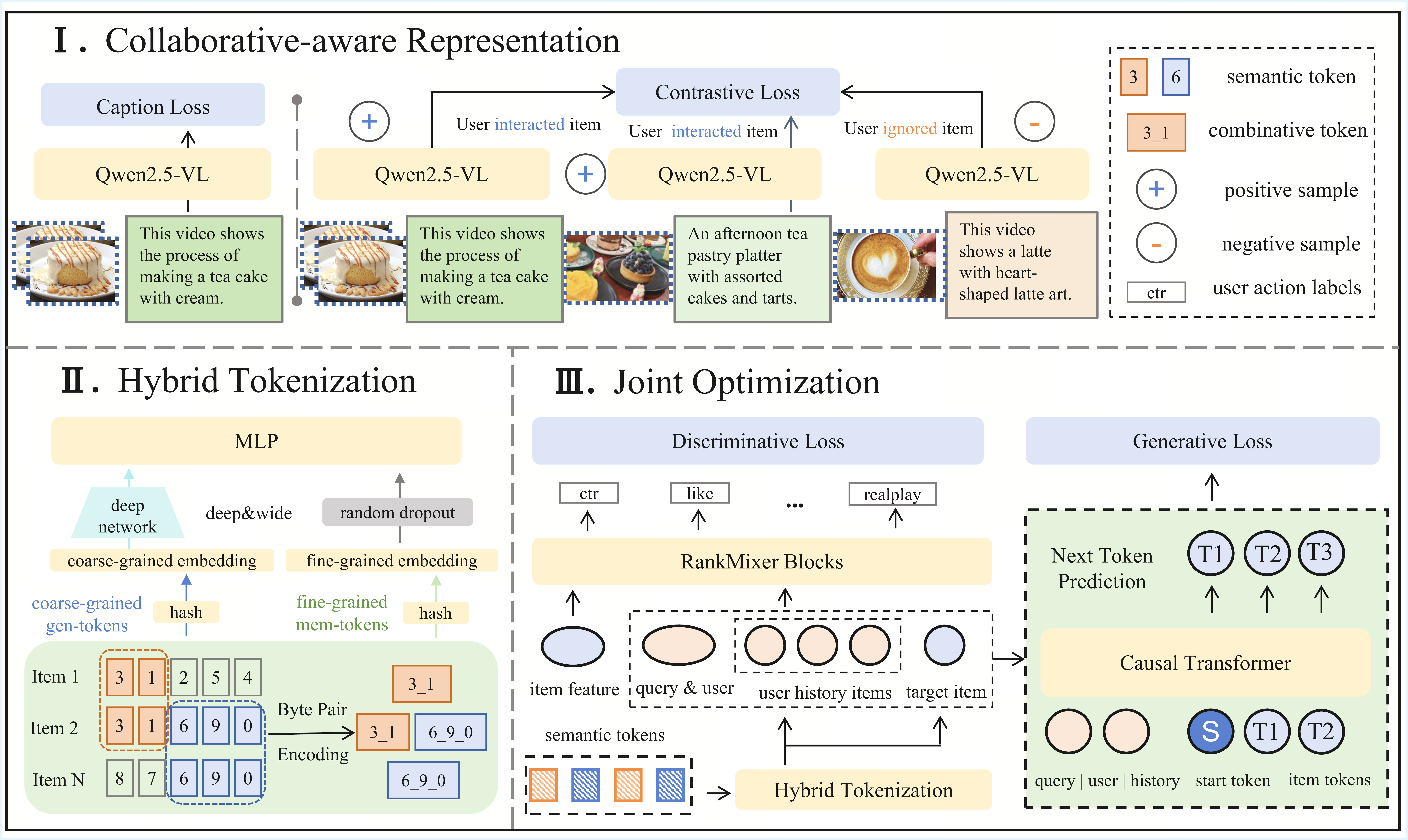}
    \caption{Framework of TRM.}
    \label{fig:pipeline}
\end{figure*}

\paragraph{Scaling laws for ranking models.}
A broad empirical study surveys scaling phenomena across diverse large recommendation backbones \citep{guo2024insights} and \cite{yan2025unlocking} provides a practical path unlocking scaling properties in production settings. 
Recent work has begun to uncover analogous behaviors in ranking and recommendation models \citep{scaling1,ardalani2022understanding}. \cite{fang2024scaling} provide evidence of power-law scaling in dense retrieval when evaluated with continuous surrogates, giving a basis for compute–data trade-off analyses. At industrial scale, generative recommenders further confirmed these trends, with OneRec reporting consistent performance gains under substantially increased training FLOPs \citep{deng2025onerec,zhou2025onerectr}, while OneSearch exploring unified end-to-end generation for e-commerce search with quantized item encodings \citep{chen2025onesearch}.

\section{Methodology}

We propose a unified tokenization-based ranking framework. As shown in Figure \ref{fig:pipeline}, the framework reformulates item modeling, representation discretization, and ranking optimization into a single pipeline.






\subsection{Collaborative-aware Multimodal Item Representation}
To learn item representations that are both semantically grounded and behaviorally aligned, we adapt the MLLM for representation learning by explicitly aligning embeddings with collaborative signals. We construct two types of training pairs from interaction logs. The first type consists of query–item pairs derived from positive user feedback. The second type consists of item–item pairs with high collaborative similarity by frequent co-clicks. We apply contrastive learning to both pair types to align representations of collaboratively similar queries and items, while separating unrelated samples. The contrastive alignment loss is defined as:
\begin{equation}
\mathcal{L}_{\text{align}}
= - \mathbb{E}_{(a,b)\in\mathcal{P}}
\log
\frac{
\exp\left(\mathrm{sim}(\mathbf{h}_a, \mathbf{h}_b)/\tau\right)
}{
\sum_{b' \in \mathcal{B}}
\exp\left(\mathrm{sim}(\mathbf{h}_a, \mathbf{h}_{b'})/\tau\right)
},
\end{equation}
where $(a,b)$ denotes a positive query–item or item–item pair, $h$ is the last token of the MLLM denoting the item representation, $\mathcal{B}$ is the set of in-batch samples, $\tau$ is a temperature hyperparameter, and $\mathrm{sim}(.,.)$ denotes cosine similarity.

\subsection{Hybrid Tokenization with Generalization–Memorization Trade-off}

\begin{figure}[t]
\centering
\begin{minipage}[t]{0.48\textwidth}
\centering
\includegraphics[width=\linewidth]{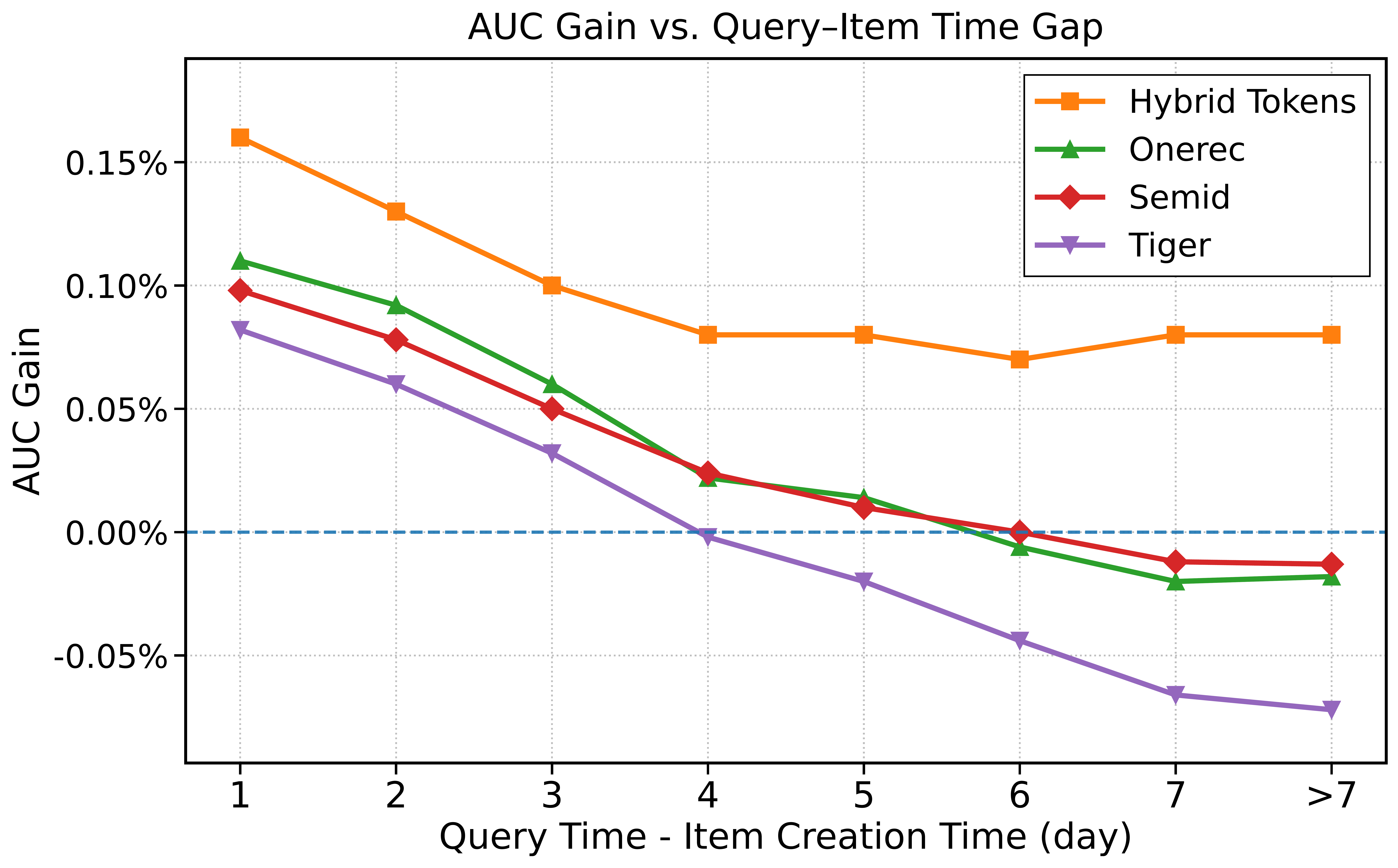}
\caption{AUC Gain compared to ID baseline by substituting different types of semantic tokens for item IDs. We report results divided by item life time when the query request happened.}
\label{fig:hybrid-compare}
\end{minipage}
\hfill
\begin{minipage}[t]{0.48\textwidth}
\centering
\includegraphics[width=\linewidth]{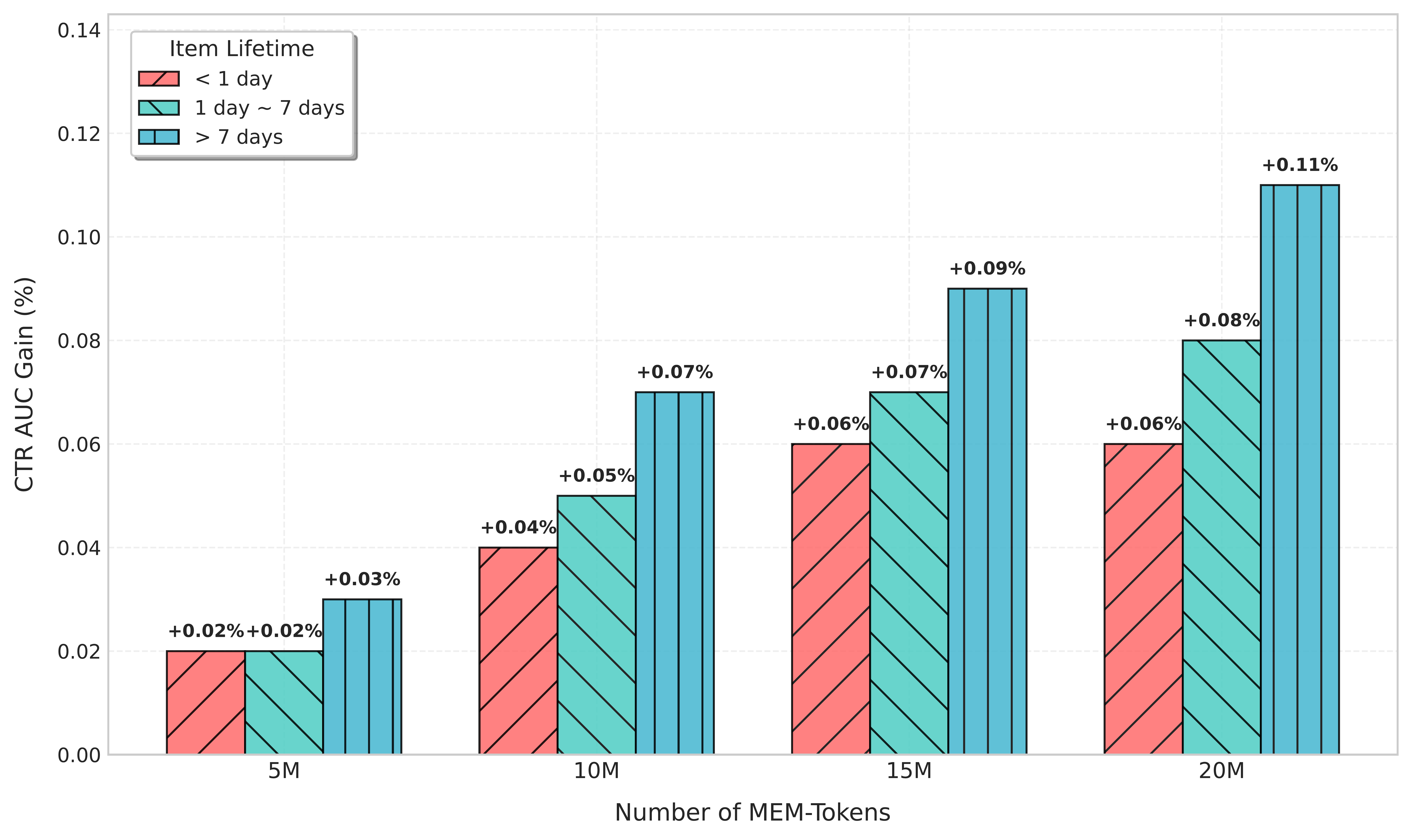}
\caption{The change of AUC gain for items of different life time when varying the number of mem-tokens. We report evaluation results of CTR AUC compared to using only gen-tokens.}
\label{fig:auc_gain_comparison}
\end{minipage}
\end{figure}

We propose hybrid tokenization that balances generalization and memorization. Here we illustrate the item-side hybird tokens and the query-side uses similar methods. Collaborative-aware item and query representations are quantized via RQ-Kmeans\cite{chen2025onesearch} into semantic tokens, which share coarse-grained knowledge and benefit new or low-exposure items (as shown in Figure \ref{fig:hybrid-compare}). However, these tokens underperform ID-based baselines overall, and performance degrades as item exposure increases, indicating a lack of item-specific knowledge. This arises because residual quantization provides only coarse clustering: 
A naïve aggregation of semantic tokens (e.g., “cake” + “candle”) lacks explicit compositional structure for meanings like “birthday”, compelling the dense network to infer such combinations internally and thus degrading its user-interest modeling capability.

To learn combinative knowledge, we introduce independent learnable tokens for high-frequency token combinations. Using Byte Pair Encoding (BPE) on semantic token sequences, we identify representative sub-word combinations and assign them as “memorization tokens” (mem-tokens), while original semantic tokens serve as “generalization tokens” (gen-tokens). Gen-tokens and mem-tokens are hashed into separate embedding modules. To balance the two, we adopt a Wide$\&$Deep network with gen-tokens on the deep side and mem-tokens on the wide side, applying random dropout to the wide side to prevent overfitting. As shown in Figure \ref{fig:hybrid-compare}, hybrid tokens enhance fine-grained knowledge preservation and consistently outperform item IDs as exposure frequency grows.

\subsection{Joint Optimization of Discriminative and Generative Objectives}

\begin{table*}[t]
\centering
\scalebox{0.8}{
\begin{tabular}{ccccccccc}
\toprule
\multicolumn{2}{c}{\multirow{2}{*}{Model}}                      & \multicolumn{2}{c}{CTR}             & \multicolumn{2}{c}{Real Play}        & \multicolumn{3}{c}{Efficiency}                                                                                                                                             \\ \cmidrule{3-9}
\multicolumn{2}{c}{}                                            & AUC              & QAUC             & AUC              & QAUC             & \begin{tabular}[c]{@{}c@{}}Dense\\ Param\end{tabular} & \begin{tabular}[c]{@{}c@{}}Sparse\\ Param\end{tabular} & \begin{tabular}[c]{@{}c@{}}FLOPs\\ (bs=4096)\end{tabular} \\ \midrule
ID Baseline                  & DLRM-MLP                         & 0.8662           & 0.8826           & 0.8413           & 0.8604           & 7M                                                   & 7.52T                                                  & 0.13T                                                     \\ \midrule
\multirow{5}{*}{ID-based}    & DCN-v2                           & +0.08\%          & +0.06\%          & +0.15\%          & +0.10\%          & 78M                                                   & 7.52T                                                  & 1.92T                                                     \\
                             & DHEN                             & +0.29\%          & +0.23\%          & +0.33\%          & +0.31\%          & 242M                                                  & 7.52T                                                  & 8.42T                                                     \\
                             & WuKong                           & +0.44\%          & +0.38\%          & +0.59\%          & +0.45\%          & 355M                                                  & 7.52T                                                  & 18.96T                                                    \\
                             & Pure Transformer                 & +0.56\%          & +0.46\%          & +0.69\%          & +0.56\%          & 326M                                                  & 7.52T                                                  & 11.05T                                                    \\
                             & RankMixer                        & +0.58\%          & +0.48\%          & +0.76\%          & +0.63\%          & 335M                                                  & 7.52T                                                  & 13.20T                                                    \\ \midrule
\multirow{5}{*}{Token-based} & Tiger-token                      & +0.45\%          & +0.40\%          & +0.58\%          & +0.45\%          & 338M                                                  & 2.17T                                                  & 13.89T                                                    \\
                             & OneRec-token                  & +0.53\%          & +0.45\%          & +0.64\%          & +0.54\%          & 342M                                                  & 2.14T                                                  & 13.98T                                                    \\
                             & SEMID                       & +0.56\%          & +0.47\%          & +0.75\%          & +0.61\%          & 345M                                                  & 2.21T                                                  & 14.23T                                                    \\ \cmidrule{2-9}
                             & \textbf{TRM-Pure Transformer} & \textbf{+0.61\%} & \textbf{+0.53\%} & \textbf{+0.81\%} & \textbf{+0.68\%} & \textbf{341M}                                         & \textbf{2.11T}                                         & \textbf{12.17T}
                                                                        \\
                               & \textbf{TRM-RankMixer}        & \textbf{+0.65\%} & \textbf{+0.54\%} & \textbf{+0.85\%} & \textbf{+0.70\%} & \textbf{352M}                                         & \textbf{2.11T}                                         & \textbf{14.66T}                                         \\

\bottomrule       
\end{tabular}
}
\caption{Main results. Token-based methods except TRM-Pure Transformer are trained with the same RankMixer architecture. For Tiger-token, OneRec-token and Semid, we use their semantic tokens to replace the item id in RankMixer.}
\label{tab:main-res}
\end{table*}

To fully exploit the structural information of the semantic tokens, we propose to jointly optimize both the discriminative object and the generative object. As shown in Figure \ref{fig:pipeline}, the model input comprises query features $X_Q$, item features $X_I$ (including hybrid tokens), and user features $X_U$. For the discriminative objective, all features are used to predict user actions (e.g., click, like) via BCE loss:
\begin{equation}
\mathcal{L}_d = \mathbb{E}_{(X_{Q,U,I}^i, Y^i)}\text{BCE}(Y^i,P(\hat{Y}|X_{Q,U,I}^i,\theta_d)),
\end{equation}
where $Y^i\in\{0,1\}$ is the action label. For the generative objective, $X_Q$ and $X_U$ are encoded into context tokens and used to autoregressively predict the item's semantic token sequence $s^i$ on positively interacted items ($Y^i=1$), optimized by next-token prediction loss:
\begin{equation}
\mathcal{L}_g =  \mathbb{E}_{(X_{Q,U}^i, Y^i)}  [Y^i=1] \cdot \sum_{j=1}^{L}\text{CE}(s_j^i, P(\hat{s}_j|X_{Q,U}^i,s^i_{<j},\theta_g)),
\end{equation}
where $s_j^i$ is the $j$-th gen-token and $L$ is the token length. A semi-causal mask is applied: context tokens are mutually visible, while subsequent tokens follow causal masking. The final objective combines both terms:
\begin{equation}
    \mathcal{L} = \mathcal{L}_d + \lambda\cdot \mathcal{L}_g,
\end{equation}
with $\lambda$ balancing the two losses.

\section{Analysis of Scaling Law}

\subsection{Empirical Scaling Law for Ranking Models}
\label{sec:empirical_scaling}

For a fixed representation
$r\in\{\mathrm{ID},\mathrm{sem},\mathrm{hyb}\}$, let
$\mathcal{L}_r(N)$ denote the test loss at dense-tower size $N$. We fit the
floor-aware empirical scaling law
\citep{hestness2017deep,kaplan2020scaling,bahri2024explaining}:
\begin{equation}
\label{eq:empirical-scaling-law}
\begin{aligned}
    \mathcal{L}_r(N)
    &\approx
    \mathcal{L}_{\infty,r}+A_rN^{-\beta_r},\\
    \mathcal{E}_r(N)
    &:=
    \mathcal{L}_r(N)-\mathcal{L}_{\infty,r}
    \approx A_rN^{-\beta_r},\\
    \log\mathcal{E}_r(N)
    &\approx
    \log A_r-\beta_r\log N,
\end{aligned}
\end{equation}
where $\mathcal{L}_{\infty,r}$ is the fitted representation-dependent floor,
$A_r>0$, and $\beta_r>0$ is the dense-tower scaling exponent. After subtracting
the fitted floor, both ID-based and token-based models exhibit an approximately
linear trend on the log--log scale over the evaluated range. We fit
Eq.~\eqref{eq:empirical-scaling-law} using held-out test loss, while
\cref{fig:scaling_qauc} presents the corresponding ranking-quality trends under
dense-parameter and compute budgets. Detailed assumptions and the treatment of
the loss floor are provided in the supplementary material.

We next compare the fitted slopes $\beta_r$ and the normalized finite-scale
intercepts $\log C_r$ of ID-based, semantic-token, and hybrid-token ranking
models.

\subsection{Improving the Scaling Slope with Semantic Tokens}
\label{sec:scaling_slope}

To interpret the difference in $\beta_r$, let $s_r$ denote the smoothness of
the target function over representation $r$, and let
$d_{\mathrm{eff},r}$ denote its effective dimension. Classical approximation
and nonparametric estimation results are governed by a
smoothness--dimension trade-off
\citep{yarotsky2017error,lu2021deep,siegel2023optimal}, which we summarize
using
\begin{equation}
\label{eq:representation-rho}
    \rho_r
    :=
    \frac{2s_r}{2s_r+d_{\mathrm{eff},r}}.
\end{equation}
The proxy $\rho_r$ increases with smoothness and decreases with effective
dimension. We use it only to interpret the empirical trend and do not assume
an exact or proportional relation between $\rho_r$ and $\beta_r$.

Item IDs form a sparse open set with independent embeddings, resulting in
limited cross-item sharing and representation drift as items are created and
retired. Semantic tokens instead reuse a compact vocabulary across related
items. This motivates the representation hypothesis
$\rho_{\mathrm{sem}}>\rho_{\mathrm{ID}}$ and, correspondingly, a more favorable
finite-range scaling exponent.

The test-loss fits yield
$\widehat{\beta}_{\mathrm{sem}}>
\widehat{\beta}_{\mathrm{ID}}$, consistent with the quality trends in
\cref{fig:scaling_qauc}: semantic-token models improve faster as dense
parameters and compute increase, whereas ID-based models exhibit earlier
diminishing returns. This ordering is an empirical result over the evaluated
range, rather than an architecture-independent identity between $\rho_r$ and
$\beta_r$.

\subsection{Finite-Scale Advantage of Hybrid Tokens}
\label{sec:scaling_intercept}

The slope measures how rapidly performance improves with dense capacity,
whereas the intercept characterizes performance at a finite operating scale.
To make the intercept independent of the unit used for $N$, we normalize
Eq.~\eqref{eq:empirical-scaling-law} at a fixed reference size $N_0$:
\begin{equation}
\label{eq:normalized-scaling-main}
\begin{aligned}
    \mathcal{E}_r(N)
    &\approx
    C_r\left(\frac{N}{N_0}\right)^{-\beta_r},
    \\[2pt]
    C_r
    &:=A_rN_0^{-\beta_r}
    \approx \mathcal{E}_r(N_0),
    \\[2pt]
    \log\mathcal{E}_r(N)
    &\approx
    \log C_r
    -\beta_r\log\left(\frac{N}{N_0}\right).
\end{aligned}
\end{equation}
Thus, $\log C_r$ is the fitted intercept at
$\log(N/N_0)=0$, and a smaller $C_r$ indicates lower finite-scale excess loss.

Although semantic tokens provide a favorable slope, a gen-token-only tower
must reconstruct frequent combination effects from individual tokens. Let
$S$ be the gen-token sequence and let $M_K=\mathcal{B}_K(S)$ denote the
BPE-derived features from $K$ selected combinations. Because $M_K$ is
deterministic and the hybrid model retains $S$,
$\mathbb{P}(Y\mid U,Q,S,M_K)=\mathbb{P}(Y\mid U,Q,S)$. BPE therefore changes
the finite-capacity parameterization rather than the information available to
the predictor.

For a token pattern $g$, define
$\phi_g(S)=\mathbf{1}\{g\text{ occurs in }S\}$. Suppose the mem-token branch
can realize an additive direction $c\phi_g(S)$. For a finite-capacity
gen-token predictor $f_{\mathrm{sem}}$, binary cross-entropy satisfies
\begin{equation}
\label{eq:main-bpe-gradient}
\left.
\frac{\partial}{\partial c}
\mathcal{R}\big(f_{\mathrm{sem}}+c\phi_g\big)
\right|_{c=0}
=
\mathbb{E}\!\left[
\big(\sigma(f_{\mathrm{sem}})-Y\big)\phi_g(S)
\right].
\end{equation}
If this quantity is nonzero, a sufficiently small step in the opposite
direction decreases the population loss. BPE can therefore reduce
$C_{\mathrm{sem}}$ when selected combinations capture residual interactions
that are inefficiently represented by the finite dense tower.

In Table \ref{abl:random hash} we use the same dense tower
and the same number of learnable sparse parameters:
$N_d^{\mathrm{BPE}}=N_d^{\mathrm{rand}}$ and
$N_s^{\mathrm{BPE}}=N_s^{\mathrm{rand}}$.
BPE outperforms random bucketing, indicating that the improvement is not
explained solely by additional embedding capacity, but also depends on how
that capacity is allocated. Frequency alone does not guarantee predictive
value; the advantage is established empirically by this matched control.

Over the evaluated range, the fitted coefficients satisfy
$\widehat C_{\mathrm{hyb}}<
\min\{\widehat C_{\mathrm{sem}},\widehat C_{\mathrm{ID}}\}$.
Together with
$\widehat{\beta}_{\mathrm{sem}}>
\widehat{\beta}_{\mathrm{ID}}$, these results separate two complementary
effects: semantic sharing improves the dense-scaling slope, while BPE-based
combination features reduce the finite-scale intercept. We do not assume that
BPE changes the asymptotic exponent or that its benefit grows monotonically
with $K$.

\section{Experiments}

\subsection{Experiment Settings}
\paragraph{Datasets and Environment.}
The offline experiments are conducted using a large scale video search-ranking dataset. The dataset describes a search engine from three aspects. (1) Items. Each item contains frames, titles and audios of the video, with personally identifiable information removed. (2) Queries and Users, which contains text content of user queries and the users' history interactions. (3) Interactions. For each query the dataset stores whether the user positively interacted to each item.
\paragraph{Compared Models.} 
We conduct a sufficient comparison between TRM and SOTA ranking models. The item ID-based methods includes DCN\cite{dcn}, DHEN\cite{dhen}, WuKong\cite{zhang2024wukong} and RankMixer\cite{rankmixer}. As for semantic token-based methods, we reproduce the semantic tokens proposed in TIGER\cite{rajput2023generative}, OneRec\cite{deng2025onerec} and SemID\cite{zheng2025semid}. All above sematic tokens are trained as the input of RankMixer, and we keep all hyper-parameters of the network consistent. We also train an MLP-based ranking baseline of 7M dense parameters for comparison.
\paragraph{Detailed Setting of TRM.}
In collaborative alignment we use 2 million item-item pairs. For hybrid tokenizatiojn, the gen-tokens are generated using RQ-Kmeans with 5 layers and 4096 embeddings for each codebook. The BPE algorithm generates at most $2\times10^7$ tokens, which is negligible compared to the $1.3\times10^{10}$-sized item ID set used in item ID-based methods. For the joint training of TRM, both of $X_Q$ and $X_U$ are projected into 2 tokens ({\it i.e.}, $N_q=N_u=2$), and we use a 4-layer transformer network for causal prediction. We set $\lambda=0.1$ to balance the joint training.
\paragraph{Evaluation Metrics.} 
We use AUC (Area Under the Curve) and QAUC (Query-level AUC) to evaluate the performance of different models. Dense/Sparse parameter number and FLOPs are used to evaluate models' efficiency. We mainly focus CTR and Real-Play in offline experiments, where Real-Play equals to 1 if the item is watched by the user for over 10 seconds. 


\begin{figure*}[t]
  \centering
  \includegraphics[width=0.95\textwidth]{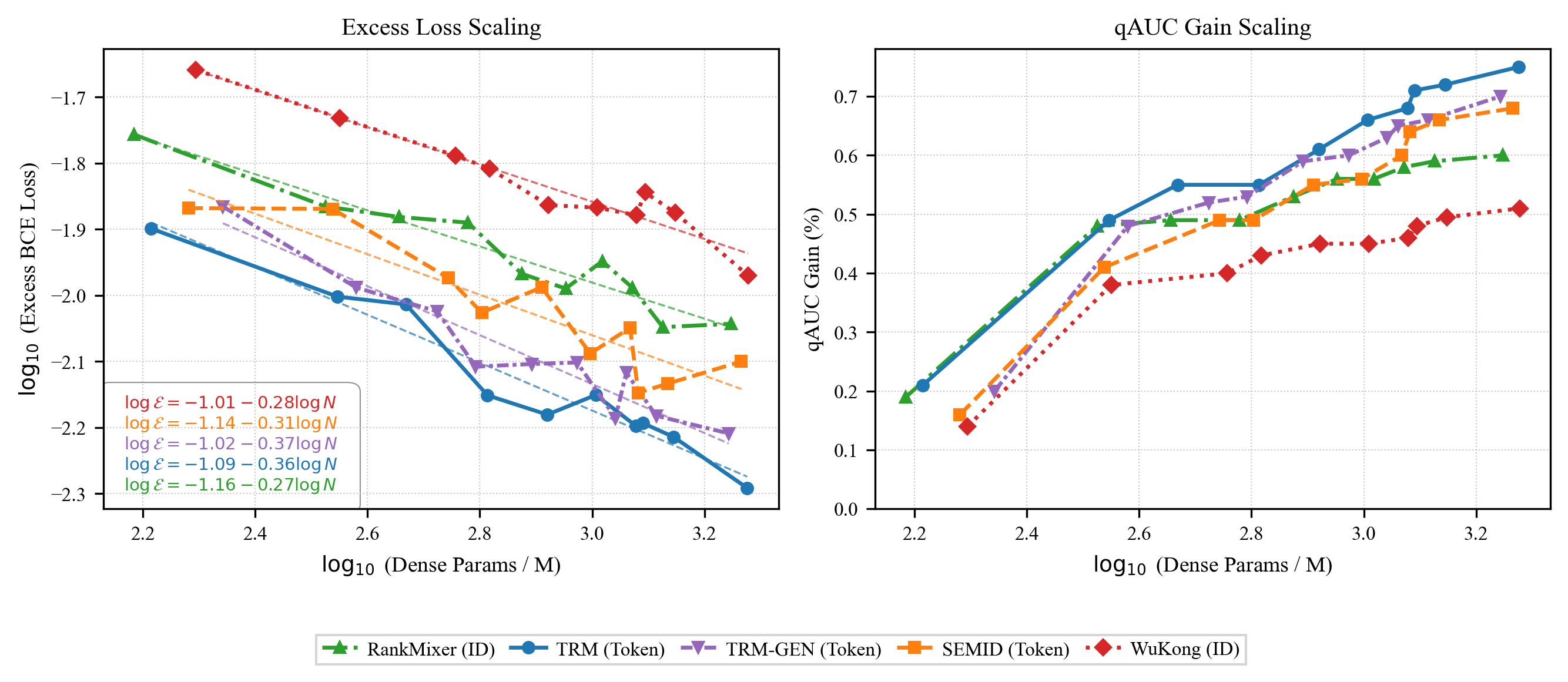}
  \caption{Test Loss (Left) and qAUC Gain (Right) comparison as dense parameters scale.}
  \label{fig:scaling_qauc}
\end{figure*}


\subsection{Main Results}

As shown in Table \ref{tab:main-res}, TRM-RankMixer maintains the best qAUC across scales, reaching +0.75\% at 1888M parameters, versus +0.68\% for SEMID and +0.60\% for ID‑based RankMixer. The widening margin indicates more effective use of added capacity. In contrast, ID‑based models saturate early: WuKong plateaus near +0.55\% at $\sim$1.9B parameters with much higher FLOPs, and DHEN saturates at only +0.42\%. This reflects the bottleneck of sparse ID memorization. Overall, \cref{fig:scaling_qauc} shows that semantic‑token rankers provide a better scaling law under both parameter and compute budgets, and TRM achieves the best quality–efficiency frontier.

\begin{table}[t]
\centering
\begin{tabular}{ccc}
\toprule
Setting          & AUC & QAUC \\ \midrule
BPE  & +0.09\%   & +0.07\%    \\
Random Hash      & +0.03\%   & +0.02\%    \\ \bottomrule
\end{tabular}
\caption{Comparison between using BPE and using Random Hash for the generation of Mem-Tokens.}
\label{abl:random hash}
\end{table}


\subsection{Scaling Law Comparison}

\label{sec:scaling_results}

\begin{table}[t]
\centering
\begin{minipage}[t]{0.48\textwidth}
\centering
\begin{tabular}{ccc}
\toprule
Setting          & AUC & QAUC \\ \midrule
w/o Col-Align    & -0.05\%   & -0.03\%    \\
w/o Hybrid Token & -0.09\%   & -0.07\%    \\
w/o Aux.NTP      & -0.05\%   & -0.05\%    \\ \bottomrule
\end{tabular}
\caption{Ablation study in different components of TRM. We report evaluation results of CTR AUC and QAUC.}
\label{abl:strategies}
\end{minipage}
\hfill
\begin{minipage}[t]{0.48\textwidth}
\centering
\scalebox{0.55}{
\begin{tabular}{c|cc|cc}
\toprule
\multirow{2}{*}{\begin{tabular}[c]{@{}c@{}}Aux Transformer\\ Setting\end{tabular}} & \multicolumn{2}{c|}{2 blocks} & \multicolumn{2}{c}{4 blocks} \\ \cmidrule{2-5} 
                                                                                  & w/ NTP loss   & w/o NTP loss  & w/ NTP loss  & w/o NTP loss  \\ \midrule
Extra Param                                                                        & +32M             & +27M             & +54M            & +48M             \\
Extra FLOPs/Batch                                                                        & +0.73T            & +0.68T             & +1.26T            & +1.19T             \\
QAUC Gain                                                                          & +0.03\%             & +0.00\%             & +0.05\%            & +0.01\%             \\ \bottomrule
\end{tabular}
}
\caption{Ablation of the generative objective. We report the evaluation results of the CTR QAUC gain.}
\label{abl:ntp}
\end{minipage}
\end{table}

As shown in \cref{fig:scaling_qauc}, when increasing the dense network capacity, TRM-RankMixer brings consistent gains and maintains the leading curve on both Params and FLOPs axes. At the largest scale in our sweep, TRM-RankMixer reaches +0.75\% qAUC gain (1888M) while SEMID and ID-based RankMixer reach +0.68\% (1843M) and +0.60\% (1768M), respectively. The increasing margin suggests that TRM enables more effective utilization of additional dense capacity, consistent with improved semantic parameter sharing and reduced dependence on sparse ID memorization.

\cref{fig:scaling_qauc} (left) shows the test loss against dense parameters $N$ on a log--log scale. All curves follow an approximately linear trend, confirming the power-law form $\log\mathcal{E}(N) \approx \log A-\beta\log N$. TRM and its variant TRM-GEN exhibit the steepest slopes ($\beta=0.36$ and $0.37$), indicating more effective utilization of added capacity. In contrast, ID-based models show flatter slopes (RankMixer: $\beta=0.27$, WuKong: $\beta=0.28$) and higher asymptotic losses, reflecting diminishing returns from sparse ID memorization. These results demonstrate that semantic-token representations yield a more favorable scaling law, consistent with the qAUC gains observed in the companion panel.


Overall, \cref{fig:scaling_qauc} demonstrates that semantic-token rankers not only improve the absolute qAUC but also provide a better scaling law under both parameter and compute budgets, and TRM consistently achieves the best quality--efficiency frontier among all compared methods.

\begin{table*}[t]
\centering
\scalebox{0.7}{
\begin{tabular}{cccccc}
\toprule
              & Search Active Days ($\uparrow$) & Change Query Ratio ($\downarrow$) & Strict CTR ($\uparrow$) & Like ($\uparrow$) & Comment ($\uparrow$) \\ \midrule
Overall       & +0.26\%         & -0.75\%                  & +0.39\%          & +1.51\%    & +1.80\%       \\ \midrule
Low-Active    & +0.27\%         & -0.81\%                  & +0.36\%          & +1.83\%    & +2.02\%       \\
Middle-Active & +0.26\%         & -0.68\%                  & +0.31\%          & +1.21\%    & +1.69\%       \\
High-Active   & +0.22\%         & -0.69\%                  & +0.40\%          & +1.20\%    & +1.59\%       \\ \bottomrule
\end{tabular}
}
\caption{Online performance of TRM. All the reported values are significant (the p value in t-test is less than 0.05).}
\label{tab:online}
\end{table*}

\begin{table}[t]
\centering
\begin{minipage}[t]{0.48\textwidth}
\centering
\begin{tabular}{ccc}
\toprule
Method           & AUC    & QAUC   \\ \midrule
BPE              & +0.09\% & +0.07\% \\
2-gram           & +0.02\% & +0.02\% \\
Prefix-ngram & +0.05\% & +0.03\% \\ \bottomrule
\end{tabular}
\caption{Performance comparison against 1-gram baseline.}
\label{tab:methods_comparison_bpe}
\end{minipage}
\hfill
\begin{minipage}[t]{0.48\textwidth}
\centering
\begin{tabular}{ccc}
\toprule
          & AUC & QAUC \\ \midrule
NTP loss & +0.05\%            & +0.05\%                              \\ 
PE & +0.02\%            & +0.01\%                            \\ 
NTP loss + PE & +0.05\%            & +0.05\%                            \\ \bottomrule
\end{tabular}
\caption{Comparison between NTP and Positional Encoding.}
\label{abl:pe}
\end{minipage}
\end{table}

\begin{table}[t]
\centering
\resizebox{0.7\linewidth}{!}{
\begin{tabular}{cccc}
\toprule
          & Item Quality & Q-I Relevance & Content Satisfactory \\ \midrule
TRM & +0.86\%            & +0.34\%             & +0.92\%                    \\ \bottomrule
\end{tabular}
}
\caption{Advantage ratio of TRM in the user review.}
\label{tab:user review}
\end{table}

\subsection{Ablation Studies}

\textbf{Effect of different strategies of TRM.} We ablate different strategies proposed in our TRM framework, as shown in Table \ref{abl:strategies}. We can see that all of the three strategies substantially improves the performance of TRM. Notably, Hybrid Tokenization contributes the most AUC gain, demonstrating the necessity of balancing the generalization and memorization capabilities of semantic tokens.

\textbf{Trade-off between generalization and memorization.} To study the impact of mem-tokens in TRM, we compare the AUC gain compared to gen-tokens when varying the number of mem-tokens. As shown in Figure \ref{fig:auc_gain_comparison}, the introduction of mem-tokens improves the performance of AUC prediction for items of all life time. We can draw two conclusions from the results. Firstly, increasing the number of mem-tokens can continuously bring about more AUC gain, but the extra gain approaches saturation when the token number increases to 20M. 
Secondly, mem-tokens contribute to memorization more than generalization. The AUC increases 0.06\% for new items ({\it i.e.}, items produced within 1 day), but the improvement comes to 0.11\% for older items ({\it i.e.}, items produced more than 7 days). This phenomenon demonstrates our insight to improve the memorization capabilities using mem-tokens.

Finally, we compare our BPE strategy with prefix-ngram which is used in SEMID\cite{zheng2025semid}. As shown in Table \ref{tab:methods_comparison_bpe}, BPE exhibits the best AUC/QAUC gain compared to 2-gram and prefix-ngram, thanks to the dynamic merging mechanism of BPE to capture combinative representations.

\textbf{Effect of the NTP loss.} Since optimization of the generative objective requires extra dense parameters and FLOPs, we ablate the NTP loss to verify the source of the QAUC gain. As shown in Table \ref{abl:ntp}, the introduction of NTP loss brings only 1.7\% dense parameter increase and 0.5\% extra FLOPs/Batch compared to the full TRM model. But removing the NTP loss would severely decrease the QAUC from +0.05\% to +0.01\%, indicating that the improvement comes from the optimization of the generative objective rather than the parameter / FLOPs increase of the extra transformer architecture. The generative objective can exploit the structural information inside the semantic token sequences. Here we compare the NTP loss with Positional Encoding (PE) which can also manually inject sequential information to semantic tokens. As shown in Table \ref{abl:pe}, the introduction of PE brings very limited performance improvement to semantic tokens. When combining NTP loss with PE, the improvement of PE even disappears. This result indicates NTP loss not only brings more sufficient structural information, but can also fully replace the role of PE.


\subsection{Online Performance}

To verify the performance of the TRM framework in real-world user-interaction scenarios, we conduct online experiments by applying the ranking model in the search engine. We report the following key performance indicators: \textit{Search Active Days} is the average number of days that each user actively uses the search engine. \textit{Change Query Ratio} is the ratio of query pages that users negatively interacted with. \textit{Strict CTR}, \textit{Like} and \textit{Comment} are the ratio of query pages that users click, like and comment in one of the shown items.

The previous baseline is a 7M DLRM ranking model. Here we replace the baseline model with TRM-Rankmixer-352M, which yields a QAUC gain of 0.54\% in CTR. From Table \ref{tab:online} we can see that TRM exhibits substantial improvement on all the indicators. Notably, TRM increases the performance across users of all activity levels, demonstrating TRM's universality in improving users' online experience. We further conduct a user review about the query pages by asking 15 real users to perform side-by-side comparison on 462 randomly sampled query pages under a double-blind setting. As shown in Table \ref{tab:user review}, TRM increases item quality, query-item relevance and content satisfactory substantially in real user feedbacks.

\section{Conclusion}

In this paper, we present TRM, a semantic token-based framework for scaling up large-scale ranking models. We perform sufficient theoretical and experimental analyses to prove that semantic token-based models demonstrates better scaling capabilities than item ID-based models. By integrating the proposed strategies ({\it i.e.}, collaborative alignment, hybrid tokenization and joint optimization), we achieves substantially better performance in the ranking system. Our online test in the large-scale search system further validates the effectiveness of TRM and reveals the promising direction of scaling up ranking models with semantic tokens.



\bibliography{aaai2027}


\newpage
\section{Technical appendices and supplementary material}
\subsection{Token-based and ID-based scaling-law analysis}
\label{app:scaling_token_id}

\subsubsection{Empirical scaling law for the tower}
\label{app:scaling-law-form}

When scaling the \emph{neural tower} while holding the representation and
training protocol fixed, we observe the standard empirical power-law behavior:
\begin{equation}
\label{eq:scaling-law-app}
\mathcal{L}(N)\;\approx\;\mathcal{L}_{\infty} \;+\; A\,N^{-\beta},
\end{equation}
where $N$ is the number of tower parameters, $\mathcal{L}_{\infty}$ is the
irreducible error level (floor), and $\beta>0$ is the fitted tower-scaling
exponent \citep{hestness2017deep,kaplan2020scaling,bahri2024explaining}.
Empirically, both ID-based and token-based models exhibit an approximately
linear trend on a log--log plot over a nontrivial range, and tokenization yields
a larger fitted exponent, i.e., a steeper tower-scaling curve
(Sec.~\ref{sec:scaling_results}).

\subsubsection{A smoothness--dimension interpretation of $\beta$}
\label{app:sd-insight}

We use a smoothness--dimension perspective to interpret the fitted exponent
$\beta$ in Eq.~\eqref{eq:scaling-law-app}. Our goal is not to derive an exact
closed-form parameter-scaling exponent from first principles. Instead, we use
classical smoothness--dimension tradeoffs as a qualitative explanation for why
the token-based tower can exhibit a larger fitted exponent than the ID-based
tower.

Let $Z\in\mathcal{Z}\subset\mathbb{R}^{d_*}$ be a continuous semantic
representation of an item, and let $(U,Q)$ denote the user/query context.
Define the Bayes-optimal CTR probability and its logit:
\[
p^*(u,q,z)=\mathbb{P}(Y=1\mid U=u,Q=q,Z=z),
\]
\[
\eta^*(u,q,z)=\log\frac{p^*(u,q,z)}{1-p^*(u,q,z)}.
\]
For fixed $(u,q)$, suppose that the target logit $\eta^*(u,q,\cdot)$ has
smoothness $s>0$ over the semantic domain. For general $s>0$, we interpret this
in the standard H\"older/Sobolev sense: higher $s$ corresponds to stronger
regularity of the target function, while larger effective dimension makes the
function class harder to approximate. In particular, when $0<s\le 1$, this
reduces to the usual zeroth-order H\"older condition
\begin{equation}
\label{eq:holder-logit-sd}
|\eta^*(u,q,z_1)-\eta^*(u,q,z_2)|
\le L_\eta \|z_1-z_2\|^{s},
\quad \forall z_1,z_2\in\mathcal{Z}.
\end{equation}
For $s>1$, we use the standard higher-order smoothness interpretation rather
than applying Eq.~\eqref{eq:holder-logit-sd} directly to function-value
differences.

Let $d_{\mathrm{eff}}$ denote an effective dimension of the semantic domain,
such as an intrinsic/manifold dimension or a data-dependent proxy for the
degrees of freedom of the representation. Approximation and nonparametric
estimation theory for H\"older/Sobolev-type function classes show that the
difficulty of learning or approximating the target is governed by the
interaction between smoothness and dimension, up to logarithmic and architecture
dependent factors \citep{yarotsky2017error,lu2021deep,siegel2023optimal}.
Motivated by this classical tradeoff, we define the following monotone
smoothness--dimension proxy:
\begin{equation}
\label{eq:rho-sd-form}
\rho(s,d_{\mathrm{eff}})
:=
\frac{2s}{2s+d_{\mathrm{eff}}}.
\end{equation}
We use $\rho(s,d_{\mathrm{eff}})$ only as a qualitative proxy: it increases with
smoothness $s$ and decreases with effective dimension $d_{\mathrm{eff}}$. Thus,
we expect a larger fitted tower-scaling exponent when the representation induces
a smoother target function over a lower-dimensional effective semantic domain.
In this sense, the empirical exponent $\beta$ is interpreted as increasing
monotonically with the smoothness--dimension favorability of the representation,
rather than being exactly equal to Eq.~\eqref{eq:rho-sd-form}.

\subsubsection{Decomposing the observed scaling: representation complexity vs.\ quantization}
\label{app:beta-decomp}

For token-based models, the tower operates on a quantized semantic variable
$\tilde Z=Q(Z)$, obtained for example by RQ-VAE or RQ-kmeans. Over the practical
tower-scaling regime, we decompose the token-based loss as
\begin{equation}
\begin{split}
\mathcal{L}_{\mathrm{tok}}(N)
&\;\approx\;
\underbrace{\mathcal{L}_{\infty}}_{\text{Bayes floor under ideal semantic $Z$}} \\
&\quad +\;
\underbrace{\Delta_{\mathrm{quant}}}_{\text{quantization-induced floor shift}} \\
&\quad +\;
\underbrace{A\,N^{-\beta_{\mathrm{tok}}}}_{\text{tower scaling term}}.
\end{split}
\label{eq:tok-scaling-decomp}
\end{equation}
Here $\Delta_{\mathrm{quant}}$ captures the information loss induced by replacing
the continuous semantic variable $Z$ with the quantized variable $\tilde Z$.
The exponent $\beta_{\mathrm{tok}}$ is the fitted exponent of the tower scaling
term after accounting for the loss floor.

This decomposition separates two effects. The representation-induced geometry
controls the complexity of the function seen by the tower, and hence the fitted
scaling behavior. Quantization, on the other hand, contributes an
$N$-independent floor shift. Therefore, after accounting for the floor, the
scaling exponent of the tower approximation term is governed primarily by the
smoothness--dimension properties of the representation. In finite-range
empirical fits, however, an unmodeled floor shift can bias the estimated slope;
therefore, $\mathcal{L}_{\infty}$ or the effective loss floor should be
estimated explicitly when fitting Eq.~\eqref{eq:scaling-law-app}.

\subsubsection{RQ-VAE quantization mainly shifts the floor}
\label{app:quant-floor-sd}

\paragraph{CTR loss is Lipschitz in logit space.}
Write binary cross-entropy in logit form:
\[
\ell(y,\eta)=\log(1+e^{\eta})-y\eta,\quad y\in\{0,1\}.
\]
Since $\frac{\partial \ell}{\partial \eta}=\sigma(\eta)-y\in[-1,1]$, we have
\begin{equation}
\label{eq:logit-lipschitz-sd}
|\ell(y,\eta)-\ell(y,\eta')|
\le |\eta-\eta'|,
\quad \forall \eta,\eta'\in\mathbb{R}.
\end{equation}

\paragraph{Quantization distortion and floor shift.}
Let $\tilde Z=Q(Z)$, and define the first-order quantization distortion
\begin{equation}
\label{eq:delta-1}
\delta_1 := \mathbb{E}\|Z-\tilde Z\|.
\end{equation}
When a squared reconstruction distortion is reported, we write
\begin{equation}
\label{eq:distortion-D}
D := \mathbb{E}\|Z-\tilde Z\|^2,
\end{equation}
so that by Jensen's inequality,
\begin{equation}
\label{eq:delta1-bound-D}
\delta_1
=
\mathbb{E}\|Z-\tilde Z\|
\le
\big(\mathbb{E}\|Z-\tilde Z\|^2\big)^{1/2}
=
D^{1/2}.
\end{equation}

Define Bayes risks with full semantic information and with quantized semantic
information:
\begin{align}
\label{eq:bayes-risks-sd}
\mathcal{L}_\infty
&:=
\inf_{h}\;
\mathbb{E}\big[\ell\big(Y, h(U,Q,Z)\big)\big],\\
\mathcal{L}_{\infty,\mathrm{tok}}
&:=
\inf_{h}\;
\mathbb{E}\big[\ell\big(Y, h(U,Q,\tilde Z)\big)\big].
\end{align}
Let
\begin{equation}
\label{eq:delta-quant-def}
\Delta_{\mathrm{quant}}
:=
\mathcal{L}_{\infty,\mathrm{tok}}-\mathcal{L}_{\infty}
\ge 0.
\end{equation}

\begin{proposition}[Quantization floor shift under CTR loss]
\label{prop:quant-floor-sd}
Assume that the Bayes logit $\eta^*$ is $L_\eta$-Lipschitz in $z$:
\begin{equation}
\label{eq:lipschitz-logit-quant}
|\eta^*(u,q,z_1)-\eta^*(u,q,z_2)|
\le
L_\eta \|z_1-z_2\|,
\quad \forall z_1,z_2\in\mathcal{Z}.
\end{equation}
Assume also that the quantized representation $\tilde Z=Q(Z)$ lies in the same
semantic domain on which $\eta^*$ is evaluated. Then
\begin{equation}
\label{eq:quant-floor-bound-sd}
0 \le \Delta_{\mathrm{quant}}
\le
L_\eta\,\mathbb{E}\|Z-\tilde Z\|
=
L_\eta\,\delta_1
\le
L_\eta\,D^{1/2}.
\end{equation}
\end{proposition}

\begin{proof}
The non-negativity follows because $(U,Q,\tilde Z)$ is a coarsening of
$(U,Q,Z)$; replacing $Z$ with $\tilde Z$ cannot reduce the Bayes risk.

For the upper bound, consider the feasible predictor for the quantized problem
defined by
\[
h_Q(u,q,\tilde z)=\eta^*(u,q,\tilde z).
\]
By the definition of $\mathcal{L}_{\infty,\mathrm{tok}}$,
\[
\mathcal{L}_{\infty,\mathrm{tok}}
\le
\mathbb{E}\big[\ell(Y,\eta^*(U,Q,\tilde Z))\big].
\]
Since the full-information Bayes risk is achieved by the Bayes logit,
\[
\mathcal{L}_{\infty}
=
\mathbb{E}\big[\ell(Y,\eta^*(U,Q,Z))\big],
\]
we obtain
\begin{align*}
\Delta_{\mathrm{quant}}
&=
\mathcal{L}_{\infty,\mathrm{tok}}-\mathcal{L}_{\infty} \\
&\le
\mathbb{E}\big[\ell(Y,\eta^*(U,Q,\tilde Z))\big]
-
\mathbb{E}\big[\ell(Y,\eta^*(U,Q,Z))\big] \\
&\le
\mathbb{E}\left[
\left|
\ell(Y,\eta^*(U,Q,\tilde Z))
-
\ell(Y,\eta^*(U,Q,Z))
\right|
\right].
\end{align*}
Using the 1-Lipschitz property of the logit-space binary cross-entropy in
Eq.~\eqref{eq:logit-lipschitz-sd}, followed by the Lipschitz continuity of
$\eta^*$ in Eq.~\eqref{eq:lipschitz-logit-quant}, gives
\[
\begin{aligned}
\Delta_{\mathrm{quant}}
&\le
\mathbb{E}\left[
|\eta^*(U,Q,\tilde Z)-\eta^*(U,Q,Z)|
\right] \\
&\le
L_\eta\,\mathbb{E}\|Z-\tilde Z\|
=
L_\eta\,\delta_1.
\end{aligned}
\]
Finally, Eq.~\eqref{eq:delta1-bound-D} gives
$\delta_1\le D^{1/2}$.
\end{proof}

\paragraph{Quantization does not change the tower exponent after accounting for the floor.}
Let $\mathcal{L}_{\mathrm{tok}}(N)$ be the best population loss achievable by an
$N$-parameter tower operating on $(U,Q,\tilde Z)$:
\[
\mathcal{L}_{\mathrm{tok}}(N)
:=
\inf_{\theta:\#\theta\le N}\;
\mathbb{E}\big[\ell(Y,g_\theta(U,Q,\tilde Z))\big].
\]
Then
\begin{equation}
\label{eq:decomp-sd}
\mathcal{L}_{\mathrm{tok}}(N)-\mathcal{L}_\infty
=
\underbrace{
\big(\mathcal{L}_{\mathrm{tok}}(N)-\mathcal{L}_{\infty,\mathrm{tok}}\big)
}_{\text{tower scaling term}}
+
\underbrace{
\Delta_{\mathrm{quant}}
}_{\text{$N$-independent floor shift}}.
\end{equation}
Therefore, if
\[
\mathcal{L}_{\mathrm{tok}}(N)-\mathcal{L}_{\infty,\mathrm{tok}}
\approx A N^{-\beta_{\mathrm{tok}}}
\]
over a practical range of $N$, then $\Delta_{\mathrm{quant}}$ contributes a
constant floor shift and does not alter the exponent of the tower scaling term
itself. However, if the floor is not modeled or if the scaling term is
comparable to the floor shift over the fitting range, the finite-range fitted
slope can be biased. This is why we fit the empirical scaling law with an
explicit floor term, as in Eq.~\eqref{eq:scaling-law-app}.

\paragraph{When can quantization be ignored?}
Quantization is negligible in the practical scaling range whenever
$\Delta_{\mathrm{quant}}$ is dominated by the tower scaling term:
\begin{equation}
\label{eq:quant-negligible-condition}
\Delta_{\mathrm{quant}}
\ll
A N^{-\beta_{\mathrm{tok}}}
\end{equation}
for the relevant values of $N$. By Proposition~\ref{prop:quant-floor-sd}, a
simple sufficient condition is
\[
L_\eta \delta_1 \ll A N^{-\beta_{\mathrm{tok}}},
\]
or, when using squared reconstruction distortion,
\[
L_\eta D^{1/2} \ll A N^{-\beta_{\mathrm{tok}}}.
\]
Thus, small quantization distortion mainly affects the asymptotic floor and has
limited impact on the fitted exponent after the floor is properly accounted for.

\subsubsection{Why tokenization can increase the observed exponent}
\label{app:why-tok-steeper}

Token-based LRMs promote \emph{semantic parameter sharing}: many items reuse a
compact set of multi-modal tokens learned from abundant head traffic, improving
transfer to tail and new items. This changes the function seen by the tower.
Instead of memorizing sparse open-set item IDs, the tower operates over a
structured semantic representation in which related items share statistical
strength through common tokens.

We hypothesize that this tokenized representation induces a more favorable
semantic geometry for tower scaling: a lower effective dimension
$d_{\mathrm{eff,tok}}$ and a smoother target function in the representation
metric. In contrast, open-set ID embeddings under long-tail frequencies and
item churn introduce weaker sharing and more representation noise, which can
increase the effective complexity seen by the tower and reduce the effective
smoothness of the target. In terms of the smoothness--dimension proxy in
Eq.~\eqref{eq:rho-sd-form}, the expected relation is
\[
\rho(s_{\mathrm{tok}},d_{\mathrm{eff,tok}})
>
\rho(s_{\mathrm{id}},d_{\mathrm{eff,id}}),
\]
which is consistent with the empirically observed ordering
\[
\beta_{\mathrm{tok}}
>
\beta_{\mathrm{id}}.
\]
Therefore, our interpretation is that tokenization increases the observed
tower-scaling exponent primarily by changing the representation-induced function
complexity, while quantization mainly contributes an $N$-independent floor shift
as analyzed above.

If representation-spectrum measurements are available, this hypothesis can be
tested by comparing the covariance-spectrum decay or effective rank of
token-based representations and ID-based embeddings. Faster spectral decay or a
smaller effective rank for token representations would support a smaller
$d_{\mathrm{eff,tok}}$. Without such a measurement, we treat spectral
concentration as a plausible mechanism rather than an independently established
empirical claim.

\end{document}